\documentclass[10pt,aps,letterpaper,nofootinbib,twocolumn,amsfonts,amsmath,amssymb,byrevtex,showpacs]{revtex4}

\usepackage[centering,margin=0.5in]{geometry}
\usepackage{bm}
\usepackage{float}
\usepackage{graphicx}
\usepackage{hyperref}

\hypersetup{%
  pdftitle={Non-conditioned generation of Schrödinger cat states in a cavity},%
  pdfauthor={Pablo Parmezani Munhoz, Antonio Vidiella-Barranco},
  pdfsubject={Quantum Optics},
  pdfkeywords={JCM, Kerr, Schrödinger cat, Q-function}}

\begin{document}

\title{Non-conditioned generation of Schr\"odinger cat states in a cavity}
    \author{P. P. Munhoz}
        \email{pmunhoz@ifi.unicamp.br}
    \author{A. Vidiella-Barranco}
        \email{vidiella@ifi.unicamp.br}
        \affiliation{Instituto de F\'\i sica `Gleb Wataghin',
        Universidade Estadual de Campinas,  13083-970 Campinas, SP,
        Brazil}
\date{\today}

\begin{abstract}

We investigate the dynamics of a two-level atom in a cavity filled
with a nonlinear medium. We show that the atom-field detuning
$\delta$ and the nonlinear parameter $\chi^{(3)}$ may be combined
to yield a periodic dynamics, allowing the generation of almost
exact superpositions of coherent states ({\sl Schr\"odinger}
cats). By analysing the atomic inversion and the field purity, we
verify that any initial atom-field state is recovered at each
revival time, and that a coherent field interacting with an
excited atom evolves to a superposition of coherent states at each
collapse time. We show that a mixed field state (statistical
mixture of two coherent states) evolves towards an almost pure
pure field state as well ({\sl Schr\"odinger} cat). We discuss the
validity of those results by using the field fidelity and the {\sl
Wigner} function.

\end{abstract}

\pacs{42.50.-p,42.50.Ct,42.50.Dv\\
Journal-ref: J.~Mod.~Opt. {\bf 52}(11), 1557 (2005)\\DOI:
10.1080/09500340500058116}

\small

\maketitle

\section{Introduction}

In quantum optics, the {\sl Jaynes}-{\sl Cummings} model ({\bf
JCM}) describes the interaction between a two-level atom and a
single quantized mode of the radiation field in a lossless cavity
and within the rotating wave approximation ({\bf RWA}). The {\bf
JCM} is probably the simplest fundamental model of field-matter
interaction with an exactly integrable Hamiltonian. Just over
forty years since its introduction \cite{jay}, the model has
originated several studies in various contexts and with different
purposes, and has become the basis for several generalizations and
other models \cite{sho}. More recently, important experimental
achievements in cavity quantum electrodynamics ({\bf QED}) and
trapped ions have stimulated both theoretical and experimental
research in that area \cite{mes}. An interesting related subject
is the quest for generation methods of macroscopically
distinguishable superpositions of quantum states, or {\sl
Schr\"odinger} cat states \cite{tri}. Several schemes using
coherent states have been proposed \cite{yur,gar,guo,dav,pat}, and
a few experimental realizations have been already accomplished in
cavity {\bf QED} as well as in trapped ions systems
\cite{bru,bru2,mon}. In cavity {\bf QED} models, states close to
those superpositions arise at specific times, for the cavity field
initially in a coherent state \cite{gea} or even in a statistical
mixture of two coherent states \cite{fre}. Propositions such as
the {\sl Yurke}-{\sl Stoler} generation scheme \cite{yur}, and
those based on quantum non-demolition processes ({\bf QND}),
depend on very large values of {\sl Kerr} nonlinearities, which is
probably the main obstacle for their implementation \cite{bru}.
However, in the last few years, the observation of large {\sl
Kerr} nonlinearities with low intensity light \cite{ima,reb,kan}
as well as propositions involving small {\sl Kerr} nonlinearities
\cite{jeo} have renewed the interest on those schemes.
Furthermore, schemes for generation involving cavity {\bf QED}
with a nonlinear medium, based on atomic conditional measurements
have also been proposed \cite{wei}.

In this paper, we present a method that does not depend on
conditional measurements. We have found that, the {\bf JCM} with a
nonlinear {\sl Kerr}-like medium, under suitable combinations of
the atom-field detuning $\delta$ and the nonlinear parameter
$\chi^{(3)}$ and for an initial field prepared either in a
coherent state or in a statistical mixture of two coherent states,
makes possible a {\sl Schr\"odinger} cat state generation with
higher fidelity than the {\bf JCM} without a nonlinear medium. We
would like to remark that in the ordinary {\bf JCM}, the initial
field in a coherent state evolves to a state close to a {\sl
Schr\"odinger} cat state at a specific time, as reported in
\cite{gea}, and if we start with a statistical mixture of two
coherent states, the field basically remains in a mixed state
\cite{fre,vid}. The possibility of generating superpositions of
coherent states in the {\bf JCM} with a nonlinear {\sl Kerr}-like
medium with an atom-field detuning has not been yet addressed in
the literature.

This paper is organized as follows: in section II we introduce the
model and obtain the evolution operator in the {\bf RWA}
approximation. In section III we present the numerical results of
some fundamental quantities and show how to obtain the condition
for a periodic dynamics. In section IV, we discuss the main
results and present our conclusions.

\section{Model}

In this section we describe the interaction of a two-level atom
with a high-Q single-mode cavity filled with a nonlinear {\sl
Kerr}-like medium, which can be modelled as an anharmonic
oscillator \cite{aga}. The cavity field is coupled with both the
two-level atom and the nonlinear medium. If the response time of
the nonlinear medium is sufficiently small we can adiabatically
eliminate the photon-photon coupling, i.e. considering the field
and nonlinear medium frequencies far from each other \cite{buz}.
Then, the total Hamiltonian of the system, with the adiabatic,
{\bf RWA} and dipole approximations, can be written as \cite{ban}
\begin{eqnarray}\label{H}
\bm{H}=\hslash\omega^{}_0\bm{a}^\dag_{}\bm{a}+\tfrac{1}{2}\hslash\omega^{}_{eg}\bm\sigma^{}_{\!z}+\hslash\chi^{(3)}_{}\bm{a}^\dag{}^2_{}\bm{a}^2_{}+\hslash\Omega(\bm{a}^\dag_{}\bm\sigma^{}_{\!-}+\bm{a}\bm\sigma^{}_{\!+}),
\end{eqnarray}
where $\omega^{}_0$ ($\omega^{}_{eg}$) is the cavity field (atomic
transition) frequency, $\bm{a}^\dag_{}$ ($\bm{a}$) is the creation
(annihilation) operator of the cavity mode obeying
$[\bm{a},\bm{a}^\dag]=\bm{1}$,
$\bm\sigma^{}_{\!z}=|e\rangle\langle{e}|-|g\rangle\langle{g}|$,
$\bm\sigma^{}_{\!+}=|e\rangle\langle{g}|$ and
$\bm\sigma^{}_{\!-}=|g\rangle\langle{e}|$ are the standard {\sl
Pauli} matrices operators, where $|e\rangle$ ($|g\rangle$) refer
to the excited (ground) atom state, $\Omega$ is the atom-field
coupling constant and $\chi^{(3)}$ is the nonlinear parameter,
proportional to the dispersive part of the third-order nonlinear
susceptibility \cite{jos}.

Following the approach of {\sl Stenholm} \cite{ste}, we delineate
the main steps to obtain the exact (under the {\bf RWA}) time
evolution operator for this model. After some algebra, we can
rewrite equation~(\ref{H}) as
\begin{eqnarray}\label{Hr}
\bm{H}=\bm{H}^{}_0+\bm{H}^{}_{\rm int},
\end{eqnarray}
where
\begin{subequations}\label{H0&Hint}
\begin{eqnarray}
&&\bm{H}^{}_0=\hslash\omega^{}_0(\bm{a}^\dag_{}\bm{a}+\tfrac{1}{2}\bm\sigma^{}_{\!z}),\\
&&\bm{H}^{}_{\rm int}=\bm{H}^{}_1+\bm{H}^{}_2,
\end{eqnarray}
\end{subequations}
with
\begin{subequations}\label{H1&H2}
\begin{eqnarray}
\hspace{-1ex}\bm{H}^{}_1=\tfrac{1}{2}\hslash\chi^{(3)}_{}+\hslash\chi^{(3)}_{}[(\bm{a}^\dag_{}\bm{a})^2_{}+\bm{a}^\dag_{}\bm{a}\bm\sigma^{}_{\!z}]-\hslash\chi^{(3)}_{}(\bm{a}^\dag_{}\bm{a}+\tfrac{1}{2}\bm\sigma^{}_{\!z}),\\
\hspace{-1ex}\bm{H}^{}_2=\hslash[\tfrac{\delta}{2}-\chi^{(3)}_{}(\bm{a}^\dag_{}\bm{a}-\tfrac{1}{2})]\bm\sigma^{}_{\!z}-\tfrac{1}{2}\hslash\chi^{(3)}_{}+\hslash\Omega(\bm{a}^\dag_{}\bm\sigma^{}_{\!-}+\bm{a}\bm\sigma^{}_{\!+}),
\end{eqnarray}
\end{subequations}
where $\delta=\omega^{}_{eg}-\omega^{}_0$ is the atom-field
detuning.

We have verified that equation~(\ref{H0&Hint}a) and
equation~(\ref{H0&Hint}b) commute, and, therefore we may write
$\bm{H}^{}_{I}=\bm{U}^{}_0(t)\bm{H}^{}_{\rm
int}\bm{U}^\dag_0(t)=\bm{H}^{}_{\rm int}$, which is just the
Hamiltonian in the interaction picture. Hence, the respective time
evolution operator is given by
\begin{eqnarray}\label{UI}
\bm{U}^{}_I(t)=\bm{U}^{}_1(t)\bm{U}^{}_2(t)=\exp\left(-\dfrac{\imath}{\hslash}\bm{H}^{}_1t\right)_{}
\exp\left(-\dfrac{\imath}{\hslash}\bm{H}^{}_2t\right)_{},
\end{eqnarray}
where the exponentials have been decoupled. After some
manipulation we obtain the following form
\begin{eqnarray}\label{UIe}
\bm{U}^{}_I(t)=\left(
\begin{array}{ccc}
\bm{E}^{}_{n+1}&0\\
0&\bm{E}^{}_{n}
\end{array}
\right)\left(
\begin{array}{ccc}
\bm{A}^{}_{n+1}&\bm{B}^\dag_{n+1}\\
-\bm{B}^{}_{n+1}&\bm{A}^\dag_n
\end{array}\right),
\end{eqnarray}
with
\begin{subequations}\label{EABCS.op}
\begin{eqnarray}
&&\bm{E}^{}_{n+1}={\rm e}^{-\imath\chi^{(3)}_{}\bm{n}^2_{}t}_{},\\
&&\bm{A}^{}_{n+1}=\cos{(\tfrac{1}{2}\bm{\Omega}^{}_{n+1}t)}-\imath\bm{\gamma}^{}_{n+1}\dfrac{\sin{(\tfrac{1}{2}\bm{\Omega}^{}_{n+1}t)}}{\bm{\Omega}^{}_{n+1}},\\
&&\bm{B}^{}_{n+1}=2\imath\Omega\bm{a}^\dag_{}\dfrac{\sin{(\tfrac{1}{2}\bm{\Omega}^{}_{n+1}t)}}{\bm{\Omega}^{}_{n+1}},
\end{eqnarray}
\end{subequations}
where
\begin{eqnarray}\label{Rabi.op}
\bm{\Omega}^{}_{n+1}=\sqrt{\bm{\gamma}^2_{n+1}+4\Omega^2_{}(\bm{n}+1)},
\end{eqnarray}
and $\bm\gamma^{}_{n+1}=\delta-2\chi^{(3)}_{}\bm{n}$.

The result in equation (\ref{Rabi.op}) has been already obtained,
in another context \cite{xie}.

\section{Atom-Field Dynamics}

In what follows we are going to assume an uncorrelated initial
atom-field state, i.e.
\begin{eqnarray}\label{rho}
\bm\rho=\bm\rho^{}_{\rm a}\otimes\bm\rho^{}_{\rm f},
\end{eqnarray}
where $\bm\rho^{}_{\rm a}$ is the initial atom density operator,
initially an excited state\footnote{The extension to a more
general initial atomic state, like $\bm\rho^{}_{\rm
a}=\sum\limits^e_{i,j=g}\rho^{}_{i,j}|i\rangle\langle{j}|$ where
$\rho^{}_{i,j}=\langle{i}|\bm\rho^{}_{\rm a}|j\rangle$, may be
easily done.} $\bm\rho^{}_{\rm a}=|e\rangle\langle{e}|$ and
$\bm\rho^{}_{\rm f}$ is the initial field density operator,
initially either a coherent state $\bm\rho^{\rm cs}_{\rm
f}=|\alpha\rangle\langle\alpha|$ or an equally weighted
statistical mixture of two coherent states $\bm\rho^{\rm sm}_{\rm
f}=\tfrac{1}{2}(|\alpha\rangle\langle\alpha|+|{-}\alpha\rangle\langle{-}\alpha|)$.
In all cases $\alpha=|\alpha|{\rm e}^{\imath\phi}_{}$, and we will
fix $|\alpha|=\sqrt{\bar{n}}=5$. The general form of the initial
field state in the {\sl Fock} state basis is
\begin{eqnarray}\label{rho.f}
\bm\rho^{}_{\rm f}=\sum\limits_{n,m}\rho^{}_{n,m}|n\rangle\langle{m}|,
\end{eqnarray}
where $\rho^{}_{n,m}=\langle{n}|\bm\rho^{}_{\rm f}|m\rangle$ are
the initial field matrix elements. For the coherent state
\begin{eqnarray}\label{rhocs}
\rho^{\rm cs}_{n,m}={\rm e}^{-|\alpha|^2_{}}_{}\dfrac{\alpha^n_{}\alpha^\ast{}^m_{}}{\sqrt{n!m!}},
\end{eqnarray}
and for the statistical mixture of two coherent states
\begin{eqnarray}\label{rhosm}
\rho^{\rm sm}_{n,m}=\tfrac{1}{2}{\rm e}^{-|\alpha|^2_{}}_{}\dfrac{\alpha^n_{}\alpha^\ast{}^m_{}}{\sqrt{n!m!}}[1+(-1)^{n+m}_{}].
\end{eqnarray}
Hence, the evolved atom-field state is given by
\begin{eqnarray}\label{rho.t}
\bm\rho(t)=\bm{U}^{}_I(t)\bm\rho\bm{U}^\dag_I(t)=\left(
\begin{array}{cc}
\bm\rho^{}_{ee}(t)&\bm\rho^{}_{eg}(t)\\
\bm\rho^{}_{ge}(t)&\bm\rho^{}_{gg}(t)
\end{array}
\right),
\end{eqnarray}
whose elements in the atomic basis are
\begin{subequations}\label{rho.t.elem.}
\begin{eqnarray}
&&\hspace{-3em}\bm\rho^{}_{ee}(t)=\sum\limits_{n,m}\rho^{}_{n,m}E^{}_{n+1}E^\ast_{m+1}A^{}_{n+1}A^\ast_{m+1}|n\rangle\langle{m}|,\\
&&\hspace{-3em}\bm\rho^{}_{eg}(t)=-\sum\limits_{n,m}\rho^{}_{n,m}E^{}_{n+1}E^\ast_{m+1}A^{}_{n+1}B^\ast_{m+1}|n\rangle\langle{m+1}|,\\
&&\hspace{-3em}\bm\rho^{}_{ge}(t)=-\sum\limits_{n,m}\rho^{}_{n,m}E^{}_{n+1}E^\ast_{m+1}B^{}_{n+1}A^\ast_{m+1}|n+1\rangle\langle{m}|,\\
&&\hspace{-3em}\bm\rho^{}_{gg}(t)=\sum\limits_{n,m}\rho^{}_{n,m}E^{}_{n+1}E^\ast_{m+1}B^{}_{n+1}B^\ast_{m+1}|n+1\rangle\langle{m+1}|,
\end{eqnarray}
\end{subequations}
with
\begin{subequations}\label{EABCS}
\begin{eqnarray}
&&E^{}_{n+1}={\rm e}^{-\imath\chi^{(3)}_{}n^2_{}t}_{},\\
&&A^{}_{n+1}=\cos{(\tfrac{1}{2}\Omega^{}_{n+1}t)}-\imath\gamma^{}_{n+1}\dfrac{\sin{(\tfrac{1}{2}\Omega^{}_{n+1}t)}}{\Omega^{}_{n+1}},\\
&&B^{}_{n+1}=2\imath\Omega\sqrt{n+1}\dfrac{\sin{(\tfrac{1}{2}\Omega^{}_{n+1}t)}}{\Omega^{}_{n+1}},
\end{eqnarray}
\end{subequations}
being
\begin{eqnarray}\label{Rabi}
\Omega^{}_{n+1}=\sqrt{\gamma^2_{n+1}+4\Omega^2_{}(n+1)},
\end{eqnarray}
the generalized {\sl Rabi} frequency, with
$\gamma^{}_{n+1}=\delta-2\chi^{(3)}_{}n$.

Once that $\bm\rho(t)$ belongs to the trace class operators acting
in the space corresponding to the direct product in
equation~(\ref{rho}) we can trace over the field variables in
equation~(\ref{rho.t}) to obtain the reduced atomic density
operator
\begin{eqnarray}\label{rho.a.t}
\bm\rho^{}_{\rm a}(t)={\rm Tr}^{}_{\rm f}[\bm\rho(t)]=\left(
\begin{array}{lll}
\lambda^{}_{ee}&\lambda^{}_{eg}\\
\lambda^\ast_{eg}&\lambda^{}_{gg}
\end{array}
\right),
\end{eqnarray}
where
$\lambda^{}_{ij}=\sum\limits_n\langle{n}|\bm\rho^{}_{ij}(t)|n\rangle$.
Using equation~(\ref{rho.t.elem.}), we have
\begin{eqnarray}\label{rho.a.t.2}
&\hspace{-3em}\nonumber\bm\rho^{}_{\rm a}(t)=\sum\limits_n\rho^{}_{n,n}|A^{}_{n+1}|^2_{}|e\rangle\langle{e}|+\sum\limits_n\rho^{}_{n,n}|B^{}_{n+1}|^2_{}|g\rangle\langle{g}|&\\
&\hspace{3em}-\sum\limits_n(\rho^{}_{n+1,n}{\rm e}^{-\imath\chi^{(3)}_{}(2n+1)t}_{}A^{}_{n+2}B^\ast_{n+1}|e\rangle\langle{g}|+{\rm c.c}).&
\end{eqnarray}
Analogously, by tracing over the atomic variables, we obtain the
(reduced) field density operator
\begin{eqnarray}\label{rho.f.t}
\bm{\rho}^{}_{\rm f}(t)={\rm Tr}^{}_{\rm a}[\bm{\rho}(t)]=\sum\limits_{n,m}\rho^{}_{n,m}(t)|n\rangle\langle{m}|,
\end{eqnarray}
where
\begin{eqnarray}\label{rho.f.t.nm}
&\hspace{-1ex}\nonumber\rho^{}_{n,m}(t)=\langle{n}|\bm{\rho}^{}_{\rm f}(t)|m\rangle=E^{}_{n+1}E^\ast_{m+1}(\rho^{}_{n,m}A^{}_{n+1}A^\ast_{m+1}&\\
&+\:\rho^{}_{n-1,m-1}{\rm e}^{2\imath\chi^{(3)}_{}(n-m)t}_{}B^{}_{n}B^\ast_{m}),&
\end{eqnarray}
are the evolved field matrix elements.

\subsection{Atomic Inversion}

A quantity usually measured in experimental cavity {\bf QED} is
the atomic population inversion \cite{bru3,pho}, defined as the
difference between the probabilities of finding the atom in the
excited state and in the ground state. Here the atomic inversion
is given by
\begin{eqnarray}\label{inversion}
{\cal W}(t)={\rm Tr}^{}_{\rm a}[\bm\sigma^{}_{\!z}\bm\rho^{}_{\rm a}(t)]=\sum\limits_nP^{}_n(|A^{}_{n+1}|^2_{}-|B^{}_{n+1}|^2_{}),
\end{eqnarray}
where $P^{}_n=\rho^{}_{n,n}=\langle{n}|\bm\rho^{}_{\rm
f}|n\rangle$ is the initial field photon number distribution.

\begin{figure}
\includegraphics[width=\linewidth]{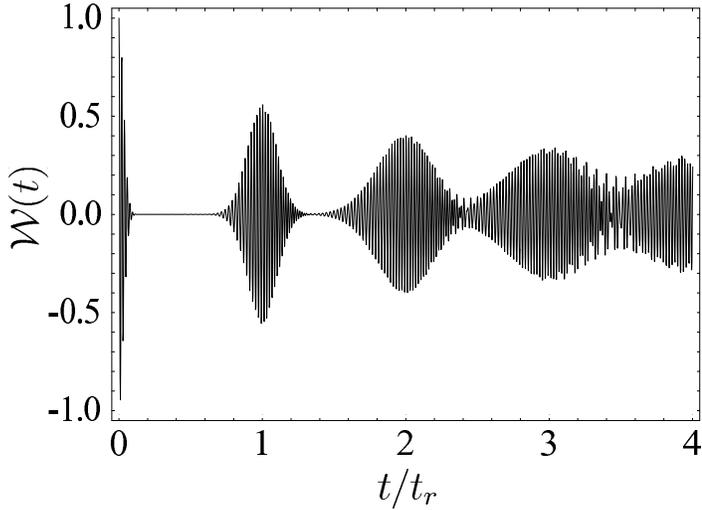}
\protect\caption{Atomic inversion as a function of $t/t^{}_r$ when
the field is initially in $\bm\rho^{\rm cs}_{\rm
f}=|\alpha\rangle\langle\alpha|$ or $\bm\rho^{\rm sm}_{\rm
f}=\frac{1}{2}(|\alpha\rangle\langle\alpha|+|{-}\alpha\rangle\langle{-}\alpha|)$
($\alpha=5$, $\phi=0$) and the atom is initially in
$\bm\rho^{}_{\rm a}=|e\rangle\langle{e}|$ with $\delta=0$ and
$\chi^{(3)}_{}=0$.} \label{figure1}
\end{figure}

It is well known that the atomic inversion is very sensitive to
the initial field photon number distribution $P^{}_n$. For the
field in the {\sl Fock} state $\bm\rho^{\rm f}_{\rm
f}=|n\rangle\langle{n}|$ we have $P^{\rm f}_{n}=\delta^{}_{n,m}$,
resulting a sinusoidal behaviour for ${\cal W}(t)$. If the initial
field is the thermal state $\bm\rho^{\rm th}_{\rm
f}=\sum\limits_nP^{\rm th}_n|n\rangle\langle{n}|$ we have $P^{\rm
th}_n=\bar{n}^n_{}/(\bar{n}+1)^{n+1}_{}$, and a more irregular
behaviour occurs \cite{pho}. For the field states considered in
this paper we have $P^{\rm cs}_n=P^{\rm sm}_n={\rm
e}^{-|\alpha|^2_{}}_{}|\alpha|^{2n}_{}/n!$ so that the atomic
response, at least regarding the atomic inversion, is the same in
either case. The atomic inversion reveals non-classical features:
the {\sl Rabi} frequency oscillations present collapses and
revivals \cite{nar}. We have parametrized the variable time as
$t/t^{}_r$ to allow a better comparison among the different plots
in terms of the revival time $t^{}_r$. In figure~\ref{figure1}, we
plot the atomic inversion as a function of $t/t^{}_r$, having
$\delta=\chi^{(3)}_{}=0$, and we observe the pattern of
oscillations characteristic of the ordinary {\bf JCM} atomic
dynamics.

\subsection{Linear Rabi Frequency}

We would like to find under which circumstances we may have a
periodic dynamics. One way of doing that is to treat the {\sl
Rabi} frequency as a continuous quantity, so that we may expand
equation~(\ref{Rabi}) around the initial mean photon number
$\bar{n}$
\begin{eqnarray}\label{Rabi.expan}
\Omega^{}_{n+1}=\sum\limits_k\left.\dfrac{1}{k!}\dfrac{\partial^k_{}\Omega^{}_{n+1}}{\partial{n}^k_{}}\right|^{}_{n=\bar{n}}(n-\bar{n})^k_{}.
\end{eqnarray}
The first term above governs the rapid oscillations in the {\sl
Rabi} frequency while the remaining terms generates the envelopes
(revivals, super-revivals and so forth). It is well-know that two
successive terms (in the discrete spectrum) of {\sl Rabi}
frequency, i.e. $\Omega^{}_{\bar{n}+1}$ and $\Omega^{}_{\bar{n}}$
have a $2\pi$ phase difference, so that the revival time is given
by
\begin{eqnarray}\label{tr}
t^{}_{r}=2\pi\dot\Omega^{-1}_{n+1}\big|^{}_{n=\bar{n}}=\pi\left|\dfrac{\Omega^{}_{\bar{n}+1}}{\Delta^{}_{\bar{n}+1}}\right|,
\end{eqnarray}
where
$\Delta^{}_{\bar{n}+1}=\Omega^2_{}-\chi^{(3)}_{}\gamma^{}_{\bar{n}+1}$.
If only the first two terms in equation~(\ref{Rabi.expan}) are
nonzero, the {\sl Rabi} frequency exhibits a perfectly periodic
behaviour \cite{koz}. This is the case, e.g. for the
intensity-dependent {\bf JCM} \cite{fre,buc}. We show that it also
may be the case for the {\bf JCM} with a {\sl Kerr}-like medium:
from the second order derivative of the {\sl Rabi} frequency,
\begin{eqnarray}\label{sec.ord.deriv.}
\ddot\Omega^{}_{n+1}\big|^{}_{n=\bar{n}}=4\left|\dfrac{\chi^{(3)}{}^2_{}\Omega^2_{\bar{n}+1}-\Delta^2_{\bar{n}+1}}{\Omega^3_{\bar{n}+1}}\right|,
\end{eqnarray}
we have $\ddot\Omega^{}_{\bar{n}+1}=0$ if
$\Delta^{}_{\bar{n}+1}=\chi^{(3)}_{}\Omega^{}_{\bar{n}+1}$ or
equivalently
\begin{eqnarray}\label{delta.c}
\delta^{}_{c}=\dfrac{\Omega^2_{}}{2\chi^{(3)}_{}}-2\chi^{(3)}_{},\qquad\mbox{(for all $n$)}. 
\end{eqnarray}

\begin{figure}
\includegraphics[width=\linewidth]{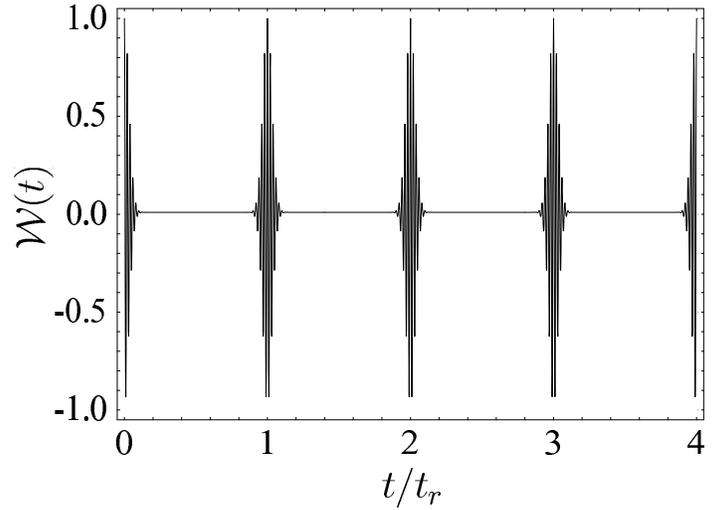}
\protect\caption{Atomic inversion as a function of $t/t^{}_r$ when
the field is initially in $\bm\rho^{\rm cs}_{\rm
f}=|\alpha\rangle\langle\alpha|$ or $\bm\rho^{\rm sm}_{\rm
f}=\frac{1}{2}(|\alpha\rangle\langle\alpha|+|{-}\alpha\rangle\langle{-}\alpha|)$
($\alpha=5$, $\phi=0$) and the atom is initially in
$\bm\rho^{}_{\rm a}=|e\rangle\langle{e}|$ with
$\delta=\delta^{}_c=4.8\Omega$ and $\chi^{(3)}_{}=0.1\Omega$.}
\label{figure2}
\end{figure}

It is clear from equation~(\ref{Rabi.expan}) that all higher-order
derivatives up to the first-order vanish when
$\delta=\delta^{}_c$. This condition determines the periodic
behaviour in the dynamics of the model. As a first illustration of
that, we plot in figure~\ref{figure2}, the atomic inversion for
the same conditions of figure~\ref{figure1}, but satisfying the
relation for $\delta^{}_c$ in equation~(\ref{delta.c}) above which
assures the periodic behaviour. Furthermore, if we insert
equation~(\ref{delta.c}) in equation~(\ref{tr}), we obtain
\begin{eqnarray}\label{tr.c}
t^{}_r=\dfrac{\pi}{\chi^{(3)}},
\end{eqnarray}
and equation~(\ref{Rabi}) becomes
\begin{eqnarray}\label{Rabi.linear}
\Omega^{}_{n+1}=\delta^{}_{c}+2\chi^{(3)}_{}(n+2).
\end{eqnarray}

We remark that similar results were obtained, in another context,
in the two-photon {\bf JCM} with {\sl Kerr}-like medium \cite{jos}
and in \cite{du}, where the authors obtained the linearized {\sl
Rabi} frequency, although they just discussed the behaviour of the
atomic inversion and used a strong-field approximation
($\bar{n}^2_{}\gg\bar{n}$) to obtain the evolution operator.

We would like now to comment about the physical relevance of the
values of $\delta$ taken in this paper, i.e. if they are
consistent with the {\bf RWA}. From experimental realizations in
microwave cavity {\bf QED} \cite{bru2,bru3,rem}, we have that
$\Omega\sim10^{4}_{}$Hz, $\omega^{}_{eg}\sim10^{6}_{}$Hz, and
$\omega^{}_{0}\sim10^{10}_{}$Hz. Here we are considering
$\delta\sim10^{2}\Omega$Hz which is consistent with the {\bf RWA}
once that $\delta\sim10^{-4}\omega_0\ll\omega_0$.

\subsection{Field Purity}

A very useful operational measure of the field state purity is
given by the linear entropy
\begin{eqnarray}\label{purity}
\zeta^{}_{\rm f}(t)=1-{\rm Tr}^{}_{\rm f}[\bm{\rho}^2_{\rm f}(t)]=1-\sum\limits_{n,m}|\rho^{}_{n,m}(t)|^2_{}.
\end{eqnarray}

\begin{figure}
\includegraphics[width=\linewidth]{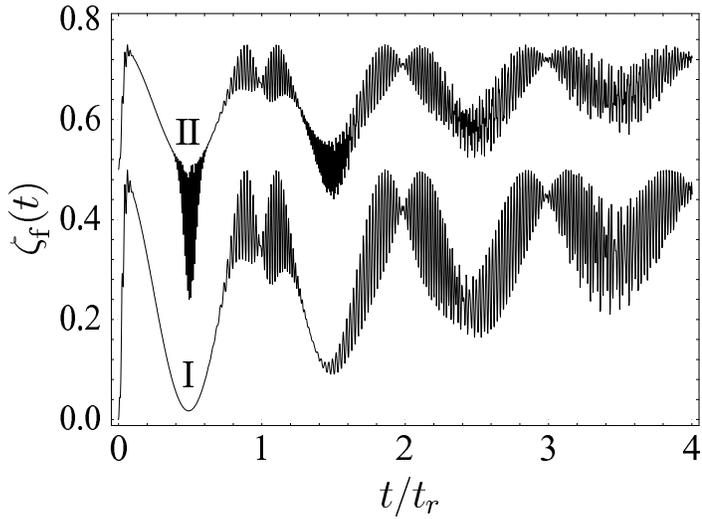}
\protect\caption{Field Purity as a function of $t/t^{}_r$ when the
field is initially in (I) $\bm\rho^{\rm cs}_{\rm
f}=|\alpha\rangle\langle\alpha|$ or (II) $\bm\rho^{\rm sm}_{\rm
f}=\frac{1}{2}(|\alpha\rangle\langle\alpha|+|{-}\alpha\rangle\langle{-}\alpha|)$
($\alpha=5$, $\phi=0$) and the atom is initially in
$\bm\rho^{}_{\rm a}=|e\rangle\langle{e}|$ with $\delta=0$ and
$\chi^{(3)}_{}=0$.} \label{figure3}
\end{figure}

In figure~\ref{figure3}, we plot the linear entropy for the
resonant case and in the absence of the {\sl Kerr}-like medium,
i.e. with $\chi^{(3)}_{}=0$. It is well-known \cite{gea} that at
half of the revival time (collapse region), the initial coherent
field evolves towards a field close to a pure ({\sl Schr\"odinger}
cat) state, figure~\ref{figure3}-I, whereas for an initial
statistical mixture of coherent states, the field is always far
from a pure state \cite{vid}, as shown in figure~\ref{figure3}-II.

\begin{figure}
\includegraphics[width=\linewidth]{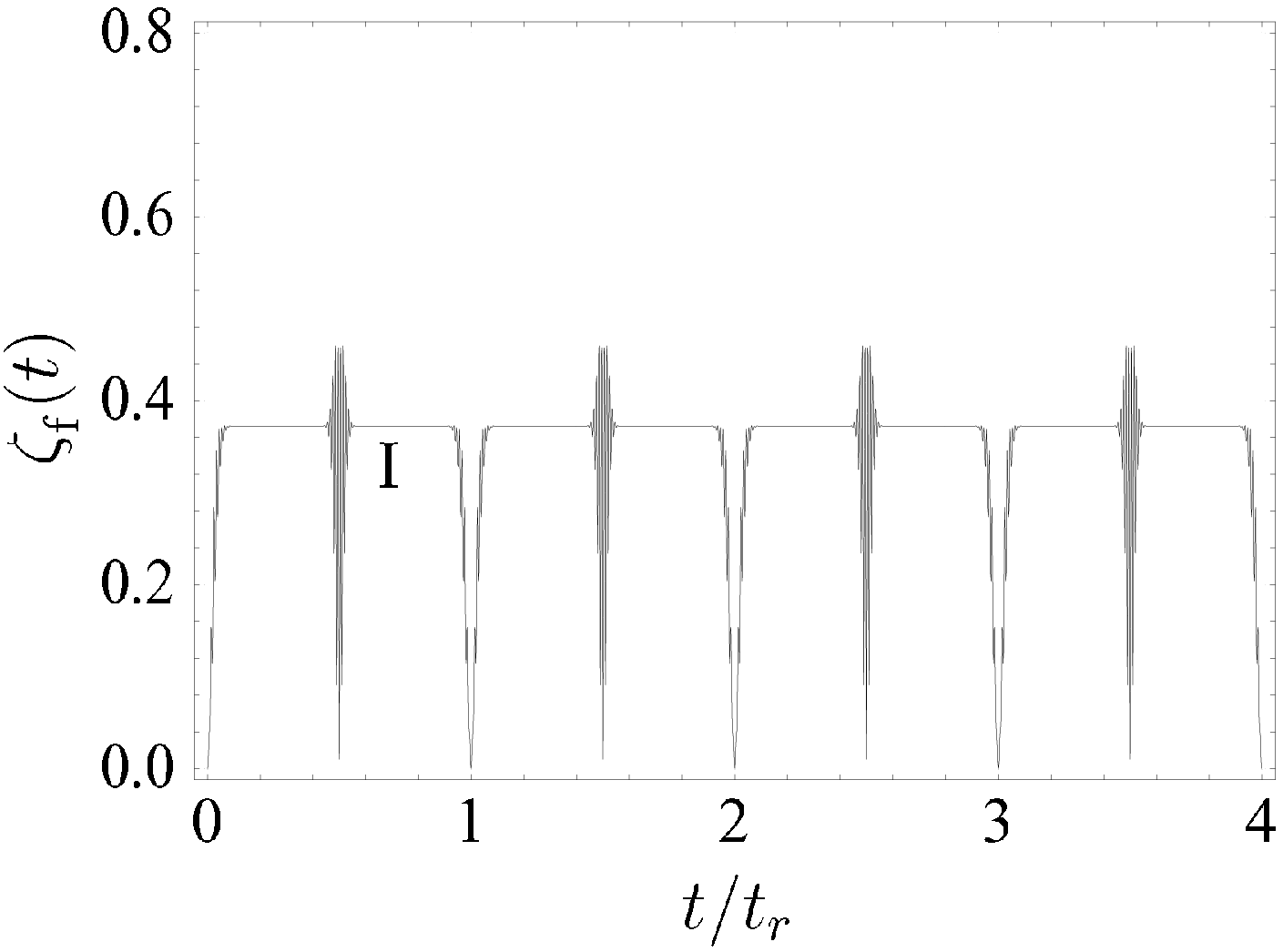}\\
\includegraphics[width=\linewidth]{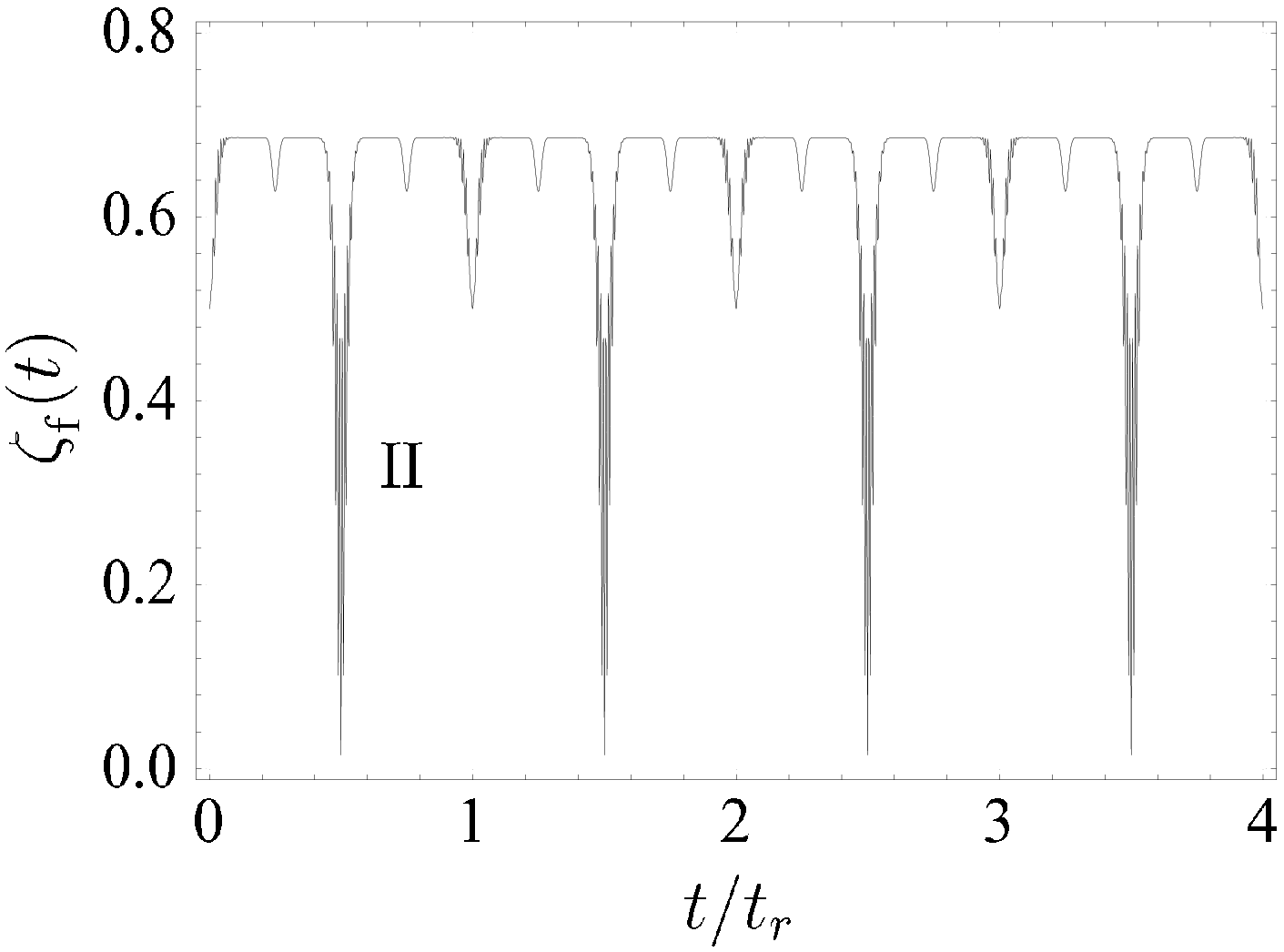}
\protect\caption{Linear entropy as a function of $t/t^{}_r$ when
the field is initially in (I) $\bm\rho^{\rm cs}_{\rm
f}=|\alpha\rangle\langle\alpha|$ or (II) $\bm\rho^{\rm sm}_{\rm
f}=\frac{1}{2}(|\alpha\rangle\langle\alpha|+|{-}\alpha\rangle\langle{-}\alpha|)$
($\alpha=5$, $\phi=0$) and the atom is initially in
$\bm\rho^{}_{\rm a}=|e\rangle\langle{e}|$ with
$\delta=\delta^{}_c=4.8\Omega$ and $\chi^{(3)}_{}=0.1\Omega$.}
\label{figure4}
\end{figure}

The situation is very different if we consider the condition that
gives a periodic dynamics. In figure~\ref{figure4} we plot the
field purity for the same conditions as considered in the
calculation of the atomic inversion. The initial coherent state
evolves to an almost pure state (very close to a superposition of
two coherent states, as we will discuss in what follows) at each
collapse time and returns to the initial state at each revival
time, figure~\ref{figure4}-I. Remarkably, an initial statistical
mixture of two coherent states also evolves to an almost pure
state (approximately a superposition of two coherent states) at
each collapse time and returns very close to the initial state at
each revival time. As seen in the atomic inversion plot in
figure~\ref{figure2}, the initial atomic state is basically
recovered at each revival time.

It is well known \cite{yur,buz2} that a nonlinear {\sl Kerr}-like
medium may convert a field in a coherent state to another pure
state, namely the {\sl Yurke}-{\sl Stoler} {\sl Schr\"odinger} cat
state. In the model present here, however, a field in an
incoherent superposition (mixed mixture) can, under suitables
conditions, evolve to a state very close to a pure state (a
coherent superposition of two coherent states). In this case of
course the other subsystem (atom) is left in a mixed state.

\subsection{Q and Wigner Functions}

In this subsection we consider the field dynamics from the point
of view of the Q and the {\sl Wigner} functions. The Q-function is
a quasi-probability distribution which is the {\sl Fourier}
transform of the anti-normally ordered quantum characteristic
function \cite{hil,cah}. For the field calculated here, the
Q-function is given by
\begin{eqnarray}\label{Q}
Q(\beta,t)=\dfrac{1}{\pi}\langle\beta|\bm{\rho}^{}_{\rm f}(t)|\beta\rangle=\dfrac{1}{\pi}{\rm e}^{-|\beta|^2_{}}_{}\sum\limits_{n,m}\rho^{}_{n,m}(t)\dfrac{\beta^{n}_{}\beta^\ast{}^m_{}}{\sqrt{n!m!}},
\end{eqnarray}
where $|\beta\rangle$ is a coherent state with
$\beta=\Re(\beta)+\imath\Im(\beta)$.

\begin{figure}
\includegraphics[width=\linewidth]{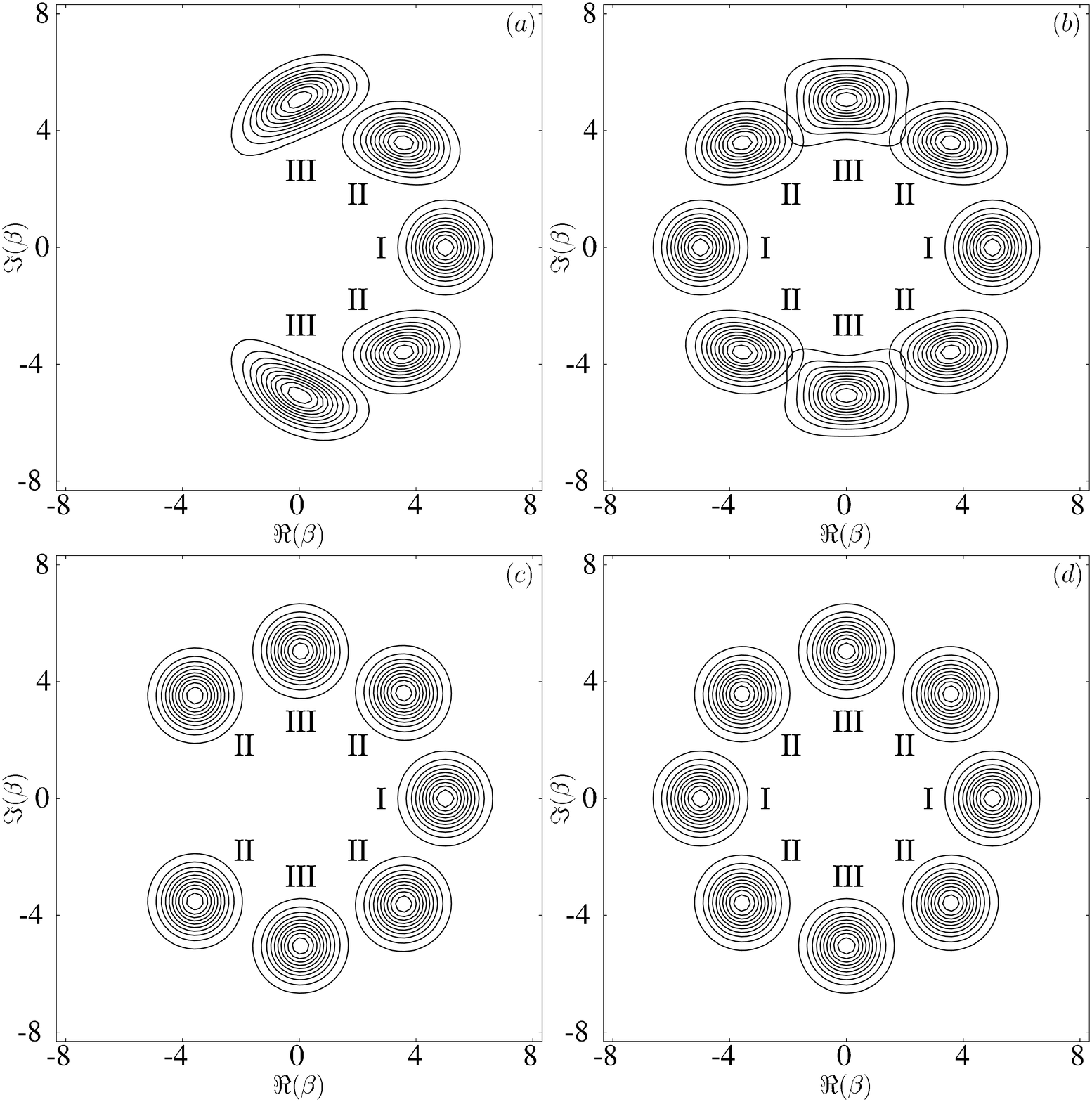}
\protect\caption{Q-function at (I) $t=0$, (II)
$t=\frac{1}{4}t^{}_r$, and (III) $t=\frac{1}{2}t^{}_r$ when the
field is initially in $(a,c)$ $\bm\rho^{\rm cs}_{\rm
f}=|\alpha\rangle\langle\alpha|$ ($\alpha=5$, $\phi=0$) or $(b,d)$
$\bm\rho^{\rm sm}_{\rm
f}=\frac{1}{2}(|\alpha\rangle\langle\alpha|+|{-}\alpha\rangle\langle{-}\alpha|)$
($\alpha=5$, $\phi=0$) and the atom is initially in
$\bm\rho^{}_{\rm a}=|e\rangle\langle{e}|$ with $(a,b)$ $\delta=0$
and $\chi^{(3)}=0$ and $(c,d)$ $\delta=\delta^{}_c=4.8\Omega$ and
$\chi^{(3)}_{}=0.1\Omega$.} \label{figure5}
\end{figure}

In discussions found in the literature, the Q-function is plotted
only at some specific times \cite{vid,jos}, as shown in
figure~\ref{figure5}. In order to provide more complete
information about the field evolution, specially at the collapse
time, when the pure state generation occurs, we present the full
time evolution of the field Q-function as a form of an animation.
The animation, external to this paper, can be
downloaded\footnote{It is necessary to download the file from {\tt
http://
journalsonline.tandf.co.uk/openurl.asp?genre=article\&id=doi:
10.1080/09500340500058116} and follow the instructions therein.}
from
\href{http://journalsonline.tandf.co.uk/openurl.asp?genre=article&id=doi:10.1080/09500340500058116}{here}.
Interesting results arise when the dynamics is periodic: there is
a relation between fractions of the revival time and the number of
peaks in which the initial coherent field splits (or recombines),
as shown in figure \ref{figure5}. For instance, at half-revival
(collapse) time, the field becomes basically a pure ({\sl
Schr\"odinger} cat) state, represented by two peaks and an
interference (oscillating) structure in phase-space. A similar
behaviour occurs if the initial field is a statistical mixture of
two coherent states: a pure {\sl Schr\"odinger} cat state
generation is almost perfect at half-revival time, although the
initial state is a mixed state.

For a better visualization of the field state generated at the
collapse time we consider the {\sl Wigner} function, a
quasi-probability distribution given by the {\sl Fourier}
transform of the symmetrically ordered characteristic function
\cite{cah}. Alternatively it may be written as \cite{moy}
\begin{eqnarray}\label{Wigner}
&\hspace{-6.4em}\nonumber{}W(\beta,t)=\dfrac{2}{\pi}\displaystyle{\sum\limits_n}(-1)^n_{}\langle{n},\beta|\bm{\rho}^{}_{\rm f}(t)|n,\beta\rangle&\\
&=\dfrac{2}{\pi}\displaystyle{\sum\limits_{n,m}}(-1)^n_{}\rho^{}_{n,m}(t)\langle{m}|\bm{D}(2\beta)|n\rangle,&
\end{eqnarray}
where $\bm{D}(2\beta)={\rm
e}^{2(\beta\bm{a}^\dag_{}-\beta^\ast_{}\bm{a})}_{}$ is the {\sl
Glauber} displacement operator and
\begin{eqnarray}\label{mDn}
&\hspace{-20em}\nonumber\langle{m}|\bm{D}(2\beta)|n\rangle=&\\&\left\{
\begin{array}{ll}
{\rm e}^{-2|\beta|^2_{}}\displaystyle{\sqrt{\frac{n!}{m!}}}(2\beta)^{m-n}_{}L^{(m-n)}_n(4|\beta|^2_{})&\quad(m\geqslant{n}),\\
{\rm e}^{-2|\beta|^2_{}}\displaystyle{\sqrt{\frac{m!}{n!}}}(-2\beta^\ast_{})^{n-m}_{}L^{(n-m)}_m(4|\beta|^2_{})&\quad(m\leqslant{n}),
\end{array}
\right.&
\end{eqnarray}
where $L^{(n-m)}_m(4|\beta|^2_{})$ are the associated {\sl
Laguerre} polynomials \cite{per}.

\begin{figure}
\includegraphics[width=\linewidth]{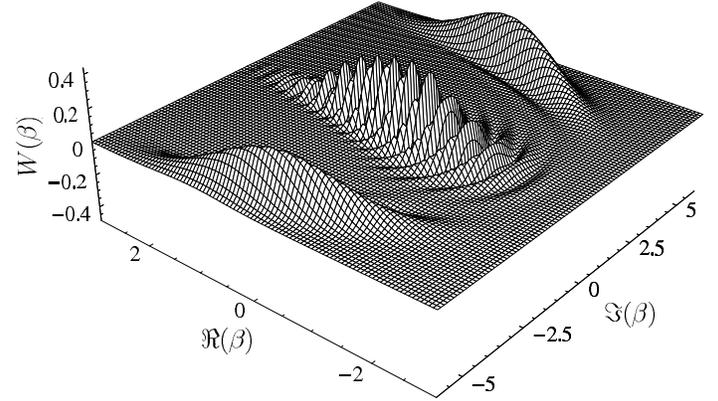}
\protect\caption{{\sl Wigner} function at $t=\frac{1}{2}t^{}_r$
when the field is initially in $\bm\rho^{\rm cs}_{\rm
f}=|\alpha\rangle\langle\alpha|$ ($\alpha=5$, $\phi=0$) and the
atom is initially in $\bm\rho^{}_{\rm a}=|e\rangle\langle{e}|$
with $\delta=0$ and $\chi^{(3)}=0$.} \label{figure6}
\end{figure}

In figure~\ref{figure6}, we plot the {\sl Wigner} function at the
collapse time for the ordinary {\bf JCM}, when the field is
initially in the coherent state and the atom is initially excited.
That corresponds to a state close to a {\sl Schr\"odinger} cat
state, in agreement with our field purity analysis,
figure~\ref{figure3}-I.

\begin{figure}
\includegraphics[width=\linewidth]{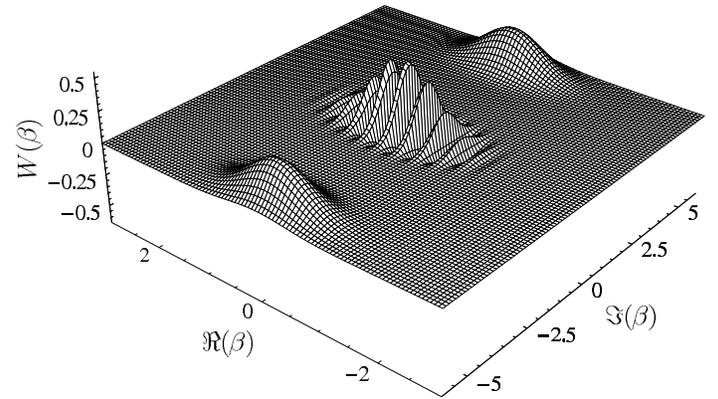}
\protect\caption{{\sl Wigner} function at $t=\frac{1}{2}t^{}_r$
when the field is initially in $\bm\rho^{\rm cs}_{\rm
f}=|\alpha\rangle\langle\alpha|$ ($\alpha=5$, $\phi=0$) and the
atom is initially in $\bm\rho^{}_{\rm a}=|e\rangle\langle{e}|$
with $\delta=\delta^{}_c=4.8\Omega$ and
$\chi^{(3)}_{}=0.1\Omega$.} \label{figure7}
\end{figure}

When the condition for periodicity is fulfilled, the {\sl Wigner}
function at the corresponding time gives the state depicted in
figure~\ref{figure7}. We clearly see that the field is much closer
to a {\sl Schr\"odinger} cat state in this case. As we shall see,
when the field is initially in a coherent state, all values of
$\delta^{}_c$ allow an almost {\sl Schr\"odinger} cat state
generation at each collapse time, with each cat state having a
specific relative phase. For complementarity, we discuss in
appendix A the situation of large detuning (dispersive
approximation) based on an effective hamiltonian \cite{pei}.

We would like to point out that the action of the {\sl Kerr}-like
medium itself is not enough to `purify' the field. The {\sl
Kerr}-like medium merely converts the statistical mixture of two
coherent states into another statistical mixture, as we show in
appendix B. From the phase space point of view, the initial
statistical mixture, figure \ref{figure5}-(b), splits in two
deformed peaks, one for each coherent state of the incoherent
superposition, in such a way that there is no perfect phase
recombination at the collapse time. In the case of periodic
dynamics, the peaks are uniform, as shown in figure
\ref{figure5}-(d), and become virtually indistinguishable at the
time they cross each other, which means that they correspond to an
almost pure state.

\subsection{Mean Photon Number}

Now we present the procedure adopted to determine the relative
phases of the superposition attained at each collapse time,
depending on the values of $\delta$ and $\chi^{(3)}_{}$ which
satisfy equation~(\ref{delta.c}). The first thing to note is that
the mean photon number at the
\begin{widetext}
\phantom{aaaaaaaaaaaaaaaaaaaaaaaaaaaaaaaaaaaaaaaaaaaaaaaaaaaaaaaaaaa}
\vspace{-4ex}
\begin{ruledtabular}
\begin{table}[H]
\begin{tabular}{ccccccccccc}
$\chi^{(3)}$ & $0$ & $0.5\Omega$ & $0.4\Omega$ & $0.3\Omega$ & $0.2\Omega$ & $0.10\bar{6}\Omega$ & $0.1\Omega$ & $0.05\Omega$ & $0.01\Omega$ & $0.005\Omega$ \\
$\delta$ & $0$ & $0$ & $0.45\Omega$ & $1.0\bar{6}\Omega$ & $2.1\Omega$ & $4.4741\bar{6}\Omega$ & $4.8\Omega$ & $9.9\Omega$ & $49.98\Omega$ & $99.99\Omega$ \\
$\bar{n}(\tfrac{1}{2}t^{}_r)$ & $25.500$ & $25.074$ & $25.110$ & $25.179$ & $25.316$ & $25.493$ & $25.495$ & $25.324$ & $25.020$ & $25.005$
\end{tabular}
\caption{Mean photon number at $t=\tfrac{1}{2}t^{}_r$ when the
field is initially in $\bm\rho^{\rm cs}_{\rm
f}=|\alpha\rangle\langle\alpha|$ or $\bm\rho^{\rm sm}_{\rm
f}=\frac{1}{2}(|\alpha\rangle\langle\alpha|+|{-}\alpha\rangle\langle{-}\alpha|)$
($\alpha=5$, $\phi=0$) and the atom is initially in
$\bm\rho^{}_{\rm a}=|e\rangle\langle{e}|$ for different values of
$\delta$ and $\chi^{(3)}_{}$.}\label{tab1}
\end{table}
\end{ruledtabular}
\end{widetext}
collapse time may not be the same as the initial one.

To verify this we consider the mean photon number given by
\begin{eqnarray}\label{mean}
\bar{n}(t)={\rm Tr}^{}_{\rm f}[\bm{\rho}^{}_{\rm f}(t)\bm{n}]=\sum\limits_nP^{}_n(t)n,
\end{eqnarray}
where $P^{}_n(t)$, for the model presented here, is given by
\begin{eqnarray}\label{Pn}
P^{}_n(t)=\rho^{}_{n,n}(t)=\langle{n}|\bm{\rho}^{}_{\rm f}(t)|n\rangle=P^{}_n|A^{}_{n+1}|^2_{}+P^{}_{n-1}|B^{}_n|^2_{}.
\end{eqnarray}

As shown in table~\ref{tab1}, for each combination of $\delta$ and
$\chi^{(3)}_{}$ we have a specific mean photon number at the
collapse time.

\subsection{Field Fidelity}

To obtain the values of the relative phase $\vartheta$, of the
superposition attained at the collapse time, we calculate the
field fidelity, defined as
\begin{eqnarray}\label{fidelity}
{\cal F}^{}_{\rm f}(t)={\rm Tr}^{}_{\rm f}[\bm{\rho}^{}_{\rm f}\bm{\rho}^{}_{\rm f}(t)]=\sum\limits_{n,m}\rho^{}_{m,n}\rho^{}_{n,m}(t), 
\end{eqnarray}
so that the evolved field state equals the initial field state if
and only if ${\cal F}^{}_{\rm f}(t)=1$. For that, we compare the
field state obtained at $t=\tfrac{1}{2}t^{}_r$, when the condition
for the periodic dynamics is satisfied, to the following ({\sl
Schr\"odinger} cat) state
\begin{eqnarray}\label{cat}
|\widetilde{\alpha};\vartheta\rangle={\cal C}^{\frac{1}{2}}_{}(|\widetilde{\alpha}\rangle+{\rm e}^{\imath\vartheta}_{}|{-}\widetilde{\alpha}\rangle), 
\end{eqnarray}
i.e. $\bm\rho^{}_{\rm f}=\bm\rho^{\rm cat}_{\rm
f}=|\widetilde{\alpha};\vartheta\rangle\langle\widetilde{\alpha};\vartheta|$,
where ${\cal C}=\tfrac{1}{2}(1+{\rm
e}^{-2|\widetilde{\alpha}|^2}_{}\cos{\vartheta})^{-1}_{}$ is the
normalization constant. From the mean photon number values at the
collapse time we have
$|\widetilde{\alpha}|^2=\bar{n}(\frac{1}{2}t_r)$ with
$\widetilde{\alpha}=|\widetilde{\alpha}|{\rm
e}^{\imath\widetilde{\phi}}_{}$, where the value
$\widetilde{\phi}=\pi/2$ was obtained from the previous analysis
of the Q-function (see figure~\ref{figure5}). We then vary the
values of the phase $\vartheta$ at the collapse time until we
obtain ${\cal F}^{}_{\rm f}(t)\approx1$. The results are presented
in table~\ref{tab2} for the initial coherent field state and the
atom excited for different values of $\delta$ and $\chi^{(3)}_{}$.
Apart from the first combination (ordinary {\bf JCM}), all the
others satisfy equation~(\ref{delta.c}) and we have the generation
of a state very close to a {\sl Schr\"odinger} cat state at each
collapse time with a specific relative phase $\vartheta$. We have
also found that, for an initial coherent field state with
$\phi=\pi$, i.e. $\bm\rho^{\rm cs}_{\rm
f}=|{-}\alpha\rangle\langle{-}\alpha|$, the relative phase of the
generated superposition is given by $-\vartheta$, instead. We have
payed special attention to the combinations (i) $\delta=4.8\Omega$
and $\chi^{(3)}_{}=0.1\Omega$, which generates a state very close
to an even coherent state ($\vartheta=0$), and (ii)
$\delta=4.4741\bar{6}\Omega$ and
$\chi^{(3)}_{}=0.10\bar{6}\Omega$,
\begin{widetext}
\phantom{aaaaaaaaaaaaaaaaaaaaaaaaaaaaaaaaaaaaaaaaaaaaaaaaaaaaaaaaaaa}
\vspace{-4ex}
\begin{ruledtabular}
\begin{table}[H]
\begin{tabular}{ccccccccccc}
$\chi^{(3)}$ & $0$ & $0.5\Omega$ & $0.4\Omega$ & $0.3\Omega$ & $0.2\Omega$ & $0.10\bar{6}\Omega$ & $0.1\Omega$ & $0.05\Omega$ & $0.01\Omega$ & $0.005\Omega$ \\
$\delta$ & $0$ & $0$ & $0.45\Omega$ & $1.0\bar{6}\Omega$ & $2.1\Omega$ & $4.4741\bar{6}\Omega$ & $4.8\Omega$ & $9.9\Omega$ & $49.98\Omega$ & $99.99\Omega$ \\
$\vartheta$ & $1.21\pi$ & $0.45\pi$ & $0.52\pi$ & $0.4\pi$ & $0.7\pi$ & $\pi$ & $0$ & $1.23\pi$ & $1.49\pi$ & $1.5\pi$ \\
${\cal F}_{\rm f}(\tfrac{1}{2}t^{}_r)$ & $0.7872$ & $0.9674$ & $0.9418$ & $0.9231$ & $0.8751$ & $0.9883$ & $0.9924$ & $0.9318$ & $0.9897$ & $0.9973$
\end{tabular}
\caption{Field fidelity at $t=\tfrac{1}{2}t^{}_r$ for the field
initially in $\bm\rho^{\rm cs}_{\rm
f}=|\alpha\rangle\langle\alpha|$ ($\alpha=5$, $\phi=0$) and atom
initially in $\bm\rho^{}_{\rm a}=|e\rangle\langle{e}|$ for
different values of $\delta$ and $\chi^{(3)}_{}$ with the
respective relative phase $\vartheta$.}\label{tab2}
\end{table}
\end{ruledtabular}
\end{widetext}
which generates a state very close to an odd coherent state
($\vartheta=\pi$): those combinations are the only ones that
enable the generation of an almost {\sl Schr\"odinger} cat state
at each collapse time when the initial field state is either a
coherent state or a statistical mixture of two coherent states. In
appendix B we (analytically) show how such results could be
understood.

\section{Conclusions}

In this work we have investigated the dynamics of a field in a
lossless cavity interacting with a two-level atom in the presence
of a nonlinear {\sl Kerr}-like medium. Using the density operator
formalism, we have obtained the exact ({\bf RWA}) evolution
operator for this model. We have found that the dynamics of the
{\bf JCM} with a {\sl Kerr}-like medium is considerably richer
than shown in the literature. The parameters $\delta$ and
$\chi^{(3)}$ may combine in a way that new and interesting
features are revealed: for instance, we may obtain a periodic
dynamics, contrarily to what happens in the ordinary {\bf JCM}.

The field interacting just with a two-level atom evolves to a
state not that close to a {\sl Schr\"odinger} cat. On the other
hand, a {\sl Kerr}-like medium alone may split a single coherent
state into a superposition of two coherent states, but a very
large nonlinearity is required for that. In any case the
generation of pure states from a mixed state is not verified. In
our model both a {\sl Kerr}-like medium and an atom conveniently
detuned from the field are important in the nonclassical field
generation process. The fine tuning of $\delta$ is important if
one wants to match a specific value of the nonlinear parameter
$\chi^{(3)}$, in order to  achieve a periodic dynamics.
Particularly, with a finite detuning there is no need of large
{\sl Kerr} nonlinearities, given that in a certain range of
parameters, the larger the detuning the smaller the nonlinearity
required. However, the detuning does not play a critical role in
the generation scheme, because even for a zero detuning we have a
value for the nonlinear parameter (large, though) that gives us a
periodic dynamics. Although schemes based on non-resonant (large
detuning) interactions do not require precise control over the
detuning, they normally rely upon conditioned measurements. In
particular, we observed that the periodic dynamics, dictated by
$\delta^{}_c$, allows us to recover the initial state at each
revival time. We have also found that an initially coherent field
becomes an almost exact superposition of coherent states ({\sl
Schr\"odinger} cat state) at each collapse time. The advantage of
our method is that generation of sharp {\sl Schr\"odinger}
cat-like states of the field may be achieved in a non-conditioned
manner, i.e. without the need of collapsing the atomic state. It
seems that the non-linearity of a {\sl Kerr}-like medium has an
effect on the field that closely resembles the non-linear
behaviour resulting from an atomic measurement, in the case of a
conditional method.

We have also found the conditions in which a field initially in a
statistical mixture of two coherent states evolves towards an
almost pure state, e.g. close to an even coherent state. Such a
quantum field `purification' may be well understood from the phase
space point of view: as the field undergoes periodic evolution, at
certain times the overlap of Q-functions coming from opposite
branches is almost perfect, meaning that an almost pure state
({\sl Schr\"odinger} cat) has been generated. In order to better
illustrate the field state generation process, we have calculated
the cavity field Q-function for successive (close enough) times in
a way that we could produce an animation of the complete evolution
of the Q-function from $t=0$ until half of the revival time. The
generation of an almost pure field state from a mixed state may of
course be achieved at the expense of the purity of the atomic
states, i.e. the atom itself ends up in a mixed state.

\section*{Acknowledgements}

We would like to thank F.~L.~Semi\~ao for a critical reading of
the manuscript, and A.~F.~Gomes, M.~A.~Marchiolli, R.~M.~Angelo,
and R.~J.~Missori for valuable suggestions and discussions. We
thank the financial support by CNPq (Conselho Nacional de
Desenvolvimento Cient\'\i fico e Tecnol\'ogico) and FAPESP
(Funda\c c\~ao de Amparo \`a Pesquisa do Estado de S\~ao Paulo).

\appendix

\section{Analytical results for the dispersive limit}

We consider here the usual procedure to obtain the dispersive
Hamiltonian for the {\bf JCM} \cite{pei}, but including a {\sl
Kerr}-like medium. The dressed states maintain their usual form
\begin{subequations}\label{eigeneq}
\begin{eqnarray}
&&|+,n\rangle=\sin\theta^{}_{n+1}|e,n\rangle+\cos\theta^{}_{n+1}|g,n+1\rangle,\\
&&|-,n\rangle={-}\cos\theta^{}_{n+1}|e,n\rangle+\sin\theta^{}_{n+1}|g,n+1\rangle,
\end{eqnarray}
\end{subequations}
but the coefficients are given by
\begin{subequations}\label{sin.cos}
\begin{eqnarray}
&&\sin\theta^{}_{n+1}=\dfrac{\omega^{}_{n+1}}{\sqrt{{(\Omega^{}_{n+1}-\gamma^{}_{n+1})}^2_{}+{\omega^2_{n+1}}}},\\
&&\cos\theta^{}_{n+1}=\dfrac{(\Omega^{}_{n+1}-\gamma^{}_{n+1})}{\sqrt{{(\Omega^{}_{n+1}-\gamma^{}_{n+1})}^2_{}+{\omega^2_{n+1}}}},
\end{eqnarray}
\end{subequations}
where $\omega^{}_{n+1}=2\Omega\sqrt{n+1}$. The corresponding
eigenvalues are
\begin{eqnarray}\label{eigenval}
E^{}_{\pm,n}=\hslash\omega^{}_0(n+\tfrac{1}{2})+\hslash\chi^{(3)}_{}n^2_{}\pm\tfrac{1}{2}\hslash\Omega^{}_{n+1}.
\end{eqnarray}
The dispersive limit is obtained when we consider $\bm{H}^{}_{\rm
int}$, equation~(\ref{H0&Hint}b), as a small perturbation of the
whole Hamiltonian \cite{pei}. It is equivalent to make
\begin{eqnarray}\label{delta.dispersive}
|\delta|\gg\omega^{}_{n+1},
\end{eqnarray}
for any `relevant' $n$\footnote{By relevant we consider the states
with significant probability $P_n=\langle{n}|\bm{\rho}^{}_{\rm
f}|n\rangle$ of population for the field under consideration.}.
Under that condition, equation~(\ref{eigenval}) becomes
\begin{eqnarray}\label{eigenval2}
E^{}_{\pm,n}\approx\hslash\omega^{}_0(n+\tfrac{1}{2})\pm\tfrac{1}{2}\hslash|\delta|+\hslash\chi^{(3)}_{}n(n\mp1)\pm\dfrac{\hslash\Omega^2_{}}{|\delta|}(n+1),
\end{eqnarray}
meaning that we can employ the following effective Hamiltonian
\begin{eqnarray}\label{H.eff}
&\nonumber\bm{H}^{\rm eff}=\hslash\omega^{}_0\bm{a}^\dag_{}\bm{a}+\tfrac{1}{2}\hslash\omega^{}_{eg}\bm\sigma^{}_{\!z}+\hslash\chi^{(3)}_{}\bm{a}^\dag{}^2_{}\bm{a}^2_{}&\\
&+\:\dfrac{\hslash\Omega^2_{}}{\delta}(\bm{a}^\dag_{}\bm{a}\bm\sigma^{}_{\!z}+\bm\sigma^{}_{\!+}\bm\sigma^{}_{\!-}).&
\end{eqnarray}
Analogously to the calculation of section II, we have the
evolution operator for the dispersive limit given by
\begin{eqnarray}\label{U.dispersive}
\bm{U}^{\rm eff}_I(t)={\rm e}^{-\imath\chi^{(3)}_{}(\bm{n}^2_{}-\bm{n})t}_{}\left( 
\begin{array}{cc}
{\rm e}^{-\imath\frac{\Omega^2_{}}{\delta}(\bm{n}+1)t}_{}&0\\
0&{\rm e}^{\imath\frac{\Omega^2_{}}{\delta}\bm{n}t}_{}
\end{array}
\right),
\end{eqnarray}
so that we may write the evolved field state as
\begin{eqnarray}\label{vector.f.t}
|\psi^{}_{\rm f}(t)\rangle=\sum\limits_nc^{}_n{\rm e}^{-\imath\chi^{(3)}_{}n^2_{}t}_{}{\rm e}^{\imath\chi^{(3)}_{}nt}_{}{\rm e}^{-\imath\frac{\Omega^2_{}}{\delta}nt}_{}|n\rangle, 
\end{eqnarray}
where $c^{}_n=\exp(-|\alpha|^2/2)\alpha^n/\sqrt{n!}$ is the
coherent state coefficient in the number state basis. We are now
able to demonstrate the initial field state being recovered at
each revival time and the generation of a {\sl Schr\"odinger} cat
state at each collapse time:

{\bf Case 1:} When $\delta=\delta^{}_c$, e.g. $\delta=49.98\Omega$
and $\chi^{(3)}_{}=0.01\Omega$, we recover the initial state at
$t_{r}=\pi/\chi^{(3)}_{}$
\begin{eqnarray}\label{recover}
|\psi^{}_{\rm f}(t^{}_r)\rangle={\rm e}^{-\frac{1}{2}|\alpha|^2_{}}_{}\sum\limits_n\dfrac{1}{\sqrt{n!}}({\rm e}^{-\imath\pi\frac{\Omega^2_{}}{\delta\chi^{(3)}_{}}}\alpha)^n|n\rangle\approx|\alpha\rangle. 
\end{eqnarray}
We remark that this result can be easily generalized for any
initial field state.

{\bf Case 2:} Similarly, at $\tfrac{1}{2}t_{r}=\pi/2\chi^{(3)}_{}$
the field evolves to
\begin{eqnarray}\label{generation.half.tr}
|\psi^{}_{\rm f}({\tfrac{1}{2}}t^{}_{r})\rangle={\rm e}^{-\frac{1}{2}|\alpha|^2_{}}_{}\sum\limits_n\dfrac{1}{\sqrt{n!}}(\imath\alpha)^n{\rm e}^{-\imath\frac{\pi}{2}{n}^2_{}}|n\rangle. 
\end{eqnarray}
If we use ${\rm
e}^{-\imath\frac{\pi}{2}n^2_{}}_{}=\tfrac{1}{2}(1+\imath)({\rm
e}^{-\imath\pi{n}}_{}-\imath)$, after multiplying by
$\tfrac{1}{\sqrt{2}}(1-\imath)$, we obtain
\begin{eqnarray}\label{generation.half.tr.rewrited}
&\nonumber|\psi^{}_{\rm f}(\tfrac{1}{2}t_{r})\rangle=\dfrac{1}{\sqrt{2}}(|{-}\imath{\rm e}^{-\imath\frac{\pi}{2}\frac{\Omega^2}{\delta\chi^{(3)}}}\alpha\rangle-\imath|\imath{\rm e}^{-\imath\frac{\pi}{2}\frac{\Omega^2}{\delta\chi^{(3)}}}\alpha\rangle)&\\
&\hspace{-3.9em}=\dfrac{1}{\sqrt{2}}(|\imath\alpha\rangle-\imath|{-}\imath\alpha\rangle),& 
\end{eqnarray}
in agreement to the numerical result ${\rm
e}^{\imath\vartheta}_{}={\rm
e}^{\imath1.49\pi}_{}\approx{-}\imath$.

\section{From a statistical mixture to the Schr\"odinger cat}
In the numerical analysis we have noted that the field initially
in a coherent state with $\phi=0$ evolves to a {\sl Schr\"odinger}
cat with $\vartheta$ and the field initially in a coherent state
with $\phi=\pi$ evolves to a {\sl Schr\"odinger} cat with
$-\vartheta$. Therefore, it is reasonable to suppose that the
field initially prepared in a statistical mixture of those two
coherent states evolves to the state
\begin{eqnarray}\label{sm.cat}
\bm{\rho}^{\rm sm\,cat}_{\rm f}=\tfrac{1}{2}(|\widetilde{\alpha};\vartheta\rangle\langle\widetilde{\alpha};\vartheta|+|\widetilde{\alpha};{-}\vartheta\rangle\langle\widetilde{\alpha};{-}\vartheta|),
\end{eqnarray}
i.e. a statistical mixture of two {\sl Schr\"odinger} cat states
with the same relative phase except for a minus sign. Finally, the
reason why only even and odd coherent states are obtained during
the evolution of an initial statistical mixture of two coherent
states becomes clear by noting that
\begin{eqnarray}\label{rho.cat}
&\hspace{-9.5em}\nonumber\bm{\rho}^{\rm cat}_{\rm f}=|\widetilde{\alpha};\vartheta\rangle\langle\widetilde{\alpha};\vartheta|={\cal C}[|\widetilde{\alpha}\rangle\langle\widetilde{\alpha}|+|{-}\widetilde{\alpha}\rangle\langle{-}\widetilde{\alpha}|&\\
&\hspace{-1em}+\:\cos{\vartheta}(|\widetilde{\alpha}\rangle\langle{-}\widetilde{\alpha}|+|{-}\widetilde{\alpha}\rangle\langle\widetilde{\alpha}|)-\imath\sin{\vartheta}(|\widetilde{\alpha}\rangle\langle{-}\widetilde{\alpha}|-|{-}\widetilde{\alpha}\rangle\langle\widetilde{\alpha}|)],&
\end{eqnarray}
and
\begin{eqnarray}\label{rho.sm.cat}
&\hspace{-9.6em}\nonumber\bm{\rho}^{\rm sm\,cat}_{\rm f}=\tfrac{1}{2}(|\widetilde{\alpha};\vartheta\rangle\langle\widetilde{\alpha};\vartheta|+|\widetilde{\alpha};{-}\vartheta\rangle\langle\widetilde{\alpha};{-}\vartheta|)&\\
&\hspace{1.5em}={\cal C}[|\widetilde{\alpha}\rangle\langle\widetilde{\alpha}|+|{-}\widetilde{\alpha}\rangle\langle{-}\widetilde{\alpha}|+\cos{\vartheta}(|\widetilde{\alpha}\rangle\langle{-}\widetilde{\alpha}|+|{-}\widetilde{\alpha}\rangle\langle\widetilde{\alpha}|)],&
\end{eqnarray}
are equal if and only if $\vartheta=k\pi$ ($k$ integer), i.e. only
for the even and odd coherent states.

\bibliography{cat}

\begin{thebibliography}{37}
\expandafter\ifx\csname natexlab\endcsname\relax\def\natexlab#1{#1}\fi
\expandafter\ifx\csname bibnamefont\endcsname\relax
  \def\bibnamefont#1{#1}\fi
\expandafter\ifx\csname bibfnamefont\endcsname\relax
  \def\bibfnamefont#1{#1}\fi
\expandafter\ifx\csname citenamefont\endcsname\relax
  \def\citenamefont#1{#1}\fi
\expandafter\ifx\csname url\endcsname\relax
  \def\url#1{\texttt{#1}}\fi
\expandafter\ifx\csname urlprefix\endcsname\relax\def\urlprefix{URL }\fi
\providecommand{\bibinfo}[2]{#2}
\providecommand{\eprint}[2][]{\url{#2}}

\bibitem{jay}
\bibinfo{author}{\bibnamefont{{Jaynes, E.~T.}}}, and
  \bibinfo{author}{\bibnamefont{{Cummings, F.~W.}}}, \bibinfo{year}{1963},
  \bibinfo{journal}{\emph{Proc.~IEEE}} \textbf{\bibinfo{volume}{51}},
  \bibinfo{pages}{89}.

\bibitem{sho}
\bibinfo{author}{\bibnamefont{{Shore, B.~W.}}}, and
  \bibinfo{author}{\bibnamefont{{Knight, P.~L.}}}, \bibinfo{year}{1993},
  \bibinfo{journal}{\emph{J.~Mod.~Opt.}} \textbf{\bibinfo{volume}{40}},
  \bibinfo{pages}{1195}.

\bibitem{mes}
\bibinfo{author}{\bibnamefont{{Messina, A.}}},
  \bibinfo{author}{\bibnamefont{{Maniscalco, S.}}}, and
  \bibinfo{author}{\bibnamefont{{Napoli, A.}}}, \bibinfo{year}{2003},
  \bibinfo{journal}{\emph{J.~Mod.~Opt.}} \textbf{\bibinfo{volume}{50}},
  \bibinfo{pages}{1}.

\bibitem{tri}
\bibinfo{author}{\bibnamefont{{Trimmer, J.~D.}}}, \bibinfo{year}{1980},
  \bibinfo{journal}{\emph{Proc.~Am.~Phys.~Soc.}} \textbf{\bibinfo{volume}{124}},
  \bibinfo{pages}{3325}.

\bibitem{yur}
\bibinfo{author}{\bibnamefont{{Yurke, B.}}}, and
  \bibinfo{author}{\bibnamefont{{Stoler, D.}}}, \bibinfo{year}{1986},
  \bibinfo{journal}{\emph{Phys.~Rev.~Lett.}} \textbf{\bibinfo{volume}{57}},
  \bibinfo{pages}{13}.

\bibitem{gar}
\bibinfo{author}{\bibnamefont{{Garraway, B.~M.}}},
  \bibinfo{author}{\bibnamefont{{Sherman, B.}}},
  \bibinfo{author}{\bibnamefont{{Moya-Cessa, H.}}},
  \bibinfo{author}{\bibnamefont{{Knight, P.~L.}}}, and
  \bibinfo{author}{\bibnamefont{{Kuriski, G.}}}, \bibinfo{year}{1994},
  \bibinfo{journal}{\emph{Phys.~Rev.~A}} \textbf{\bibinfo{volume}{49}},
  \bibinfo{pages}{535}.

\bibitem{guo}
\bibinfo{author}{\bibnamefont{{Guo, G.~C.}}}, and
  \bibinfo{author}{\bibnamefont{{Zheng, S.~B.}}}, \bibinfo{year}{1996},
  \bibinfo{journal}{\emph{Phys.~Lett.~A}} \textbf{\bibinfo{volume}{223}},
  \bibinfo{pages}{332}.

\bibitem{dav}
\bibinfo{author}{\bibnamefont{{Davidovich, L.}}},
  \bibinfo{author}{\bibnamefont{{Brune, M.}}},
  \bibinfo{author}{\bibnamefont{{Raimond, J.~M.}}}, and
  \bibinfo{author}{\bibnamefont{{Haroche, S.}}}, \bibinfo{year}{1996},
  \bibinfo{journal}{\emph{Phys.~Rev.~A}} \textbf{\bibinfo{volume}{53}},
  \bibinfo{pages}{1295}.
  
\bibitem{pat}
\bibinfo{author}{\bibnamefont{{Paternostro, M.}}},
  \bibinfo{author}{\bibnamefont{{Kim, M.~S.}}}, and
  \bibinfo{author}{\bibnamefont{{Ham, B.~S.}}}, \bibinfo{year}{2003},
  \bibinfo{journal}{\emph{Phys.~Rev.~A}} \textbf{\bibinfo{volume}{67}},
  \bibinfo{pages}{023811}.  

\bibitem{bru}
\bibinfo{author}{\bibnamefont{{Brune, M.}}},
  \bibinfo{author}{\bibnamefont{{Haroche, S.}}},
  \bibinfo{author}{\bibnamefont{{Raimond, J.~M.}}},
  \bibinfo{author}{\bibnamefont{{Davidovich, L.}}}, and
  \bibinfo{author}{\bibnamefont{{Zagury, N.}}}, \bibinfo{year}{1992},
  \bibinfo{journal}{\emph{Phys.~Rev.~A}} \textbf{\bibinfo{volume}{45}},
  \bibinfo{pages}{5193}.
  
\bibitem{bru2}
\bibinfo{author}{\bibnamefont{{Brune, M.}}},
  \bibinfo{author}{\bibnamefont{{Hagley, E.}}},
  \bibinfo{author}{\bibnamefont{{Dreyer, J.}}},
  \bibinfo{author}{\bibnamefont{{Ma\^itre, X.}}},
  \bibinfo{author}{\bibnamefont{{Maali, A.}}},
  \bibinfo{author}{\bibnamefont{{Wunderlich, C.}}},
  \bibinfo{author}{\bibnamefont{{Raimond, J.~M.}}}, and
  \bibinfo{author}{\bibnamefont{{Haroche, S.}}},
  \bibinfo{year}{1996}, \bibinfo{journal}{\emph{Phys.~Rev.~Lett.}}
  \textbf{\bibinfo{volume}{77}}, \bibinfo{pages}{4887}.  

\bibitem{mon}
\bibinfo{author}{\bibnamefont{{Monroe, C.}}},
  \bibinfo{author}{\bibnamefont{{Meekof, D.~M.}}},
  \bibinfo{author}{\bibnamefont{{King, B.~E.}}}, and
  \bibinfo{author}{\bibnamefont{{Wineland, D.~J.}}}, \bibinfo{year}{1996},
  \bibinfo{journal}{\emph{Science}} \textbf{\bibinfo{volume}{272}},
  \bibinfo{pages}{1131}.
  
\bibitem{gea}
\bibinfo{author}{\bibnamefont{{Gea-Banacloche, J.}}}, \bibinfo{year}{1987},
  \bibinfo{journal}{\emph{Phys.~Rev.~Lett.}} \textbf{\bibinfo{volume}{65}},
  \bibinfo{pages}{3385}.
  
\bibitem{fre}
\bibinfo{author}{\bibnamefont{{Freitas, D.~S.}}},
  \bibinfo{author}{\bibnamefont{{Vidiella-Barranco, A.}}}, and
  \bibinfo{author}{\bibnamefont{{Roversi, J.~A.}}}, \bibinfo{year}{1998},
  \bibinfo{journal}{\emph{Phys.~Lett.~A}} \textbf{\bibinfo{volume}{249}},
  \bibinfo{pages}{275}.
  
\bibitem{ima}
\bibinfo{author}{\bibnamefont{{Imamo\=glu, A.}}},
  \bibinfo{author}{\bibnamefont{{Schmidt, H.}}},
  \bibinfo{author}{\bibnamefont{{Woods, G.}}}, and
  \bibinfo{author}{\bibnamefont{{Deutsch, M.}}}, \bibinfo{year}{1997},
  \bibinfo{journal}{\emph{Phys.~Rev.~Lett.}} \textbf{\bibinfo{volume}{79}},
  \bibinfo{pages}{1467}.
  
\bibitem{reb}
\bibinfo{author}{\bibnamefont{{Rebi\'c, S.}}},
  \bibinfo{author}{\bibnamefont{{Tan, S.~M.}}},
  \bibinfo{author}{\bibnamefont{{Parkins, A.~S.}}}, and
  \bibinfo{author}{\bibnamefont{{Walls, D.~F.}}}, \bibinfo{year}{1999},
  \bibinfo{journal}{\emph{J.~Opt.~B}} \textbf{\bibinfo{volume}{1}},
  \bibinfo{pages}{490}.
          
\bibitem{kan}
\bibinfo{author}{\bibnamefont{{Kang, H.}}}, and
  \bibinfo{author}{\bibnamefont{{Zhu, Y.}}}, \bibinfo{year}{2003},
  \bibinfo{journal}{\emph{Phys.~Rev.~Lett.}} \textbf{\bibinfo{volume}{91}},
  \bibinfo{pages}{093601}.

\bibitem{jeo}
\bibinfo{author}{\bibnamefont{{Jeong, H.}}},
  \bibinfo{author}{\bibnamefont{{Kim, M.~S.}}},
  \bibinfo{author}{\bibnamefont{{Ralph, T.~C.}}}, and
  \bibinfo{author}{\bibnamefont{{Ham, B.~S.}}}, \bibinfo{year}{2004}, 
  \bibinfo{journal}{\emph{Phys.~Rev.~A}} \textbf{\bibinfo{volume}{70}},
  \bibinfo{pages}{061801}. 
  
\bibitem{wei}
\bibinfo{author}{\bibnamefont{{Wei, W.}}}, and
  \bibinfo{author}{\bibnamefont{{Guo, G.~C.}}}, \bibinfo{year}{1998},
  \bibinfo{journal}{\emph{Acta~Phys.~Sin.}} \textbf{\bibinfo{volume}{7}},
  \bibinfo{pages}{174}.
  
\bibitem{vid}
\bibinfo{author}{\bibnamefont{{Vidiella-Barranco, A.}}},
  \bibinfo{author}{\bibnamefont{{Moya-Cessa, H.}}}, and
  \bibinfo{author}{\bibnamefont{{Bu\v zek, V.}}}, \bibinfo{year}{1992},
  \bibinfo{journal}{\emph{J.~Mod.~Opt.}} \textbf{\bibinfo{volume}{39}},
  \bibinfo{pages}{1441}.
  
\bibitem{aga}
\bibinfo{author}{\bibnamefont{{Agarwal, G.~S.}}}, and
  \bibinfo{author}{\bibnamefont{{Puri, R.~R.}}}, \bibinfo{year}{1989},
  \bibinfo{journal}{\emph{Phys.~Rev.~A}} \textbf{\bibinfo{volume}{39}},
  \bibinfo{pages}{2969}.        

\bibitem{buz}
\bibinfo{author}{\bibnamefont{{Bu\v zek, V.}}}, and
  \bibinfo{author}{\bibnamefont{{Jex, I.}}}, \bibinfo{year}{1990},
  \bibinfo{journal}{\emph{Opt.~Commun.}} \textbf{\bibinfo{volume}{78}},
  \bibinfo{pages}{425}.

\bibitem{buz2}
\bibinfo{author}{\bibnamefont{{Bu\v zek, V.}}}, and
  \bibinfo{author}{\bibnamefont{{Vidiella-Barranco, A.}}}, and 
  \bibinfo{author}{\bibnamefont{{Knight, P.~L.}}}, \bibinfo{year}{1992},
  \bibinfo{journal}{\emph{Phys.~Rev.~A}} \textbf{\bibinfo{volume}{45}},
  \bibinfo{pages}{6570}.

\bibitem{ban}
\bibinfo{author}{\bibnamefont{{Bandyopadhyay, A.}}}, and
  \bibinfo{author}{\bibnamefont{{Gangopadhyay, G.}}}, \bibinfo{year}{1996},
  \bibinfo{journal}{\emph{J.~Mod.~Opt.}} \textbf{\bibinfo{volume}{43}},
  \bibinfo{pages}{487}.

\bibitem{jos}
\bibinfo{author}{\bibnamefont{{Joshi, A.}}}, and
  \bibinfo{author}{\bibnamefont{{Puri, R.~R.}}}, \bibinfo{year}{1992},
  \bibinfo{journal}{\emph{Phys.~Rev.~A}} \textbf{\bibinfo{volume}{45}},
  \bibinfo{pages}{5056}.
  
\bibitem{ste}
\bibinfo{author}{\bibnamefont{{Stenholm, S.}}}, \bibinfo{year}{1973},
  \bibinfo{journal}{\emph{Phys.~Rep.~C}} \textbf{\bibinfo{volume}{6}},
  \bibinfo{pages}{1}.
  
\bibitem{xie}
\bibinfo{author}{\bibnamefont{{Xie, R.~H.}}},
  \bibinfo{author}{\bibnamefont{{Xu, G.~O.}}}, and
  \bibinfo{author}{\bibnamefont{{Liu, D.~H.}}}, \bibinfo{year}{1995},
  \bibinfo{journal}{\emph{Aust.~J.~Phys.}} \textbf{\bibinfo{volume}{48}},
  \bibinfo{pages}{907}.    

\bibitem{bru3}
\bibinfo{author}{\bibnamefont{{Brune, M.}}},
  \bibinfo{author}{\bibnamefont{{Schmidt-Kaler, F.}}},
  \bibinfo{author}{\bibnamefont{{Maali, A.}}},
  \bibinfo{author}{\bibnamefont{{Dreyer, J.}}},
  \bibinfo{author}{\bibnamefont{{Hagley, E.}}},
  \bibinfo{author}{\bibnamefont{{Raimond, J.~M.}}}, and
  \bibinfo{author}{\bibnamefont{{Haroche, S.}}},
    \bibinfo{year}{1996}, \bibinfo{journal}{\emph{Phys.~Rev.~Lett.}}
  \textbf{\bibinfo{volume}{76}}, \bibinfo{pages}{1800}.

\bibitem{pho}
\bibinfo{author}{\bibnamefont{{Phoenix, S.~J.~D.}}}, and
  \bibinfo{author}{\bibnamefont{{Knight, P.~L.}}}, \bibinfo{year}{1988},
  \bibinfo{journal}{\emph{Ann.~Phys.}} \textbf{\bibinfo{volume}{186}},
  \bibinfo{pages}{381}.
  
\bibitem{nar}
\bibinfo{author}{\bibnamefont{{Narozhny, N.~B.}}},
  \bibinfo{author}{\bibnamefont{{S\'anchez-Mondrag\'on, J.~J.}}}, and
  \bibinfo{author}{\bibnamefont{{Eberly, J.~H.}}}, \bibinfo{year}{1981},
  \bibinfo{journal}{\emph{Phys.~Rev.~A}} \textbf{\bibinfo{volume}{23}},
  \bibinfo{pages}{236}.
  
\bibitem{koz}
\bibinfo{author}{\bibnamefont{{Kozierowski, M.}}}, \bibinfo{year}{2001},
  \bibinfo{journal}{\emph{J.~Mod.~Opt.}} \textbf{\bibinfo{volume}{48}},
  \bibinfo{pages}{773}.
 
\bibitem{buc}
\bibinfo{author}{\bibnamefont{{Buck, B.}}}, and
  \bibinfo{author}{\bibnamefont{{Sukumar, C.~V.}}}, \bibinfo{year}{1981},
  \bibinfo{journal}{\emph{Phys.~Lett.~A}} \textbf{\bibinfo{volume}{81}},
  \bibinfo{pages}{132}.

\bibitem{du}
\bibinfo{author}{\bibnamefont{{Du, S.~D.}}},
  \bibinfo{author}{\bibnamefont{{Gong, S.~Q.}}},
  \bibinfo{author}{\bibnamefont{{Xu, Z.~Z.}}}, and
  \bibinfo{author}{\bibnamefont{{Gong, C.~D.}}}, \bibinfo{year}{1997},
  \bibinfo{journal}{\emph{Quantum~Semiclass.~Opt.}} \textbf{\bibinfo{volume}{9}},
  \bibinfo{pages}{941}.
  
\bibitem{rem}
\bibinfo{author}{\bibnamefont{{Rempe, G.}}},
  \bibinfo{author}{\bibnamefont{{Walther, H.}}}, and
  \bibinfo{author}{\bibnamefont{{Klein, N.}}}, \bibinfo{year}{1987},
  \bibinfo{journal}{\emph{Phys.~Rev.~Lett.}} \textbf{\bibinfo{volume}{58}},
  \bibinfo{pages}{353}.
  
\bibitem{hil}
\bibinfo{author}{\bibnamefont{{Hillery, M.}}},
  \bibinfo{author}{\bibnamefont{{O'Connell, R.~F.}}},
  \bibinfo{author}{\bibnamefont{{Scully ,M.~O.}}}, and
  \bibinfo{author}{\bibnamefont{{Wigner, E.~P.}}}, \bibinfo{year}{1984},
  \bibinfo{journal}{\emph{Phys.~Rep.}} \textbf{\bibinfo{volume}{106}},
  \bibinfo{pages}{121}.    

\bibitem{cah}
\bibinfo{author}{\bibnamefont{{Cahill, K.~E.}}}, and
  \bibinfo{author}{\bibnamefont{{Glauber, R.~J.}}}, \bibinfo{year}{1969},
  \bibinfo{journal}{\emph{Phys.~Rev.}} \textbf{\bibinfo{volume}{177}},
  \bibinfo{pages}{1882}.

\bibitem{moy}
\bibinfo{author}{\bibnamefont{{Moya-Cessa, H.}}}, and
  \bibinfo{author}{\bibnamefont{{Knight, P.~L.}}}, \bibinfo{year}{1993},
  \bibinfo{journal}{\emph{Phys.~Rev.~A}} \textbf{\bibinfo{volume}{48}},
  \bibinfo{pages}{2479}.

\bibitem{per}
\bibinfo{author}{\bibnamefont{{Perelomov, A.}}}, \bibinfo{year}{1986},
  \emph{\bibinfo{title}{{Generalized Coherent States and Their Applications}}}
  (\bibinfo{publisher}{Berlim: Springer-Verlag}), p. 35.

\bibitem{pei}
\bibinfo{author}{\bibnamefont{{Peixoto~de~Faria, J.~G.}}}, and
  \bibinfo{author}{\bibnamefont{{Nemes, M.~C.}}}, \bibinfo{year}{1999},
  \bibinfo{journal}{\emph{Phys.~Rev.~A}} \textbf{\bibinfo{volume}{59}},
  \bibinfo{pages}{3918}.

\end{thebibliography}

\end{document}